\documentclass[doublecol,a4]{epl2} 
\usepackage{amsmath,amssymb}
\usepackage{graphicx}
\def\vec#1{\mbox{\boldmath $#1$}}

\newcommand{\deri}[2]{{\displaystyle \frac{\partial #1 }{\partial #2 }}}
\newcommand{\derri}[2]{{\displaystyle \frac{\partial^2 #1 }{\partial #2 }}}

\title{
Elastic heterogeneity, vibrational states, and thermal conductivity across an amorphisation transition}

\author{Hideyuki Mizuno\inst{1}\thanks{E-mail:
\email{Hideyuki.Mizuno@ujf-grenoble.fr}}
\and Stefano Mossa\inst{2}\thanks{E-mail:
\email{stefano.mossa@cea.fr}}
\and Jean-Louis Barrat\inst{1}\thanks{E-mail:
\email{jean-louis.barrat@ujf-grenoble.fr}}
}

\shorttitle{Elastic heterogeneity across an amorphisation transition}
\shortauthor{H. Mizuno, S. Mossa, and J.-L. Barrat}

\institute{
  \inst{1} Laboratory for Interdisciplinary Physics, UMR 5588, Universit\'e Grenoble 1 and CNRS, 38402 Saint Martin d'H\`eres, France \\
  \inst{2} INAC/SPrAM (UMR 5819 UJF, CNRS, CEA), CEA-Grenoble, 17 Rue des Martyrs, 38054 Grenoble, France
}


\abstract{
Disordered solids exhibit unusual properties of their vibrational states and thermal conductivities. Recent progresses have well established the concept of ``elastic heterogeneity", i.e., disordered materials show spatially inhomogeneous elastic moduli. In this study, by using molecular-dynamics simulations, we gradually introduce ``disorder" into a numerical system to control its modulus heterogeneity. The system starts from a perfect crystalline state, progressively transforms into an increasingly disordered crystalline state, and finally undergoes structural amorphisation. We monitor independently the elastic heterogeneity, the vibrational states, and the thermal conductivity across this transition, and show that the heterogeneity in elastic moduli is well correlated to vibrational and thermal anomalies of the disordered system.
}

\begin{document}
\maketitle

%%%%%%%%%%%%%%%%%%%%%%%%%%%%%%%%%%%%%%%%%%%%%%%%%%%%%%%%%%%%%%%%%%%%%%%%%%%%%%%%%%%%%%%%%%%%%%%%%%%%%%%%%%%%%%%%%%
\section{Introduction}
Disordered solids, including structural glasses, disordered crystals, or even crystals with complex unit cells, have unusual vibrational and thermal properties~\cite{elliott_1984,phillips_1981} compared to those of the corresponding crystalline materials, which are well understood by the Debye model. The density of states $g(\omega)$ shows an excess over the Debye prediction $g(\omega) \sim \omega^2$, often referred to as the  ``Boson peak" in glasses, in a frequency regime of $\omega \sim 1 [THz]$. The excess in $g(\omega)$ is mirrored by an excess in the specific heat $C(T)$ over the prediction $C \sim T^3$.
The $T$-dependence of the thermal conductivity $\kappa(T)$ is also very different from that of crystals,  with  a characteristic ``plateau" around $T \sim 10 [K]$.

The origin of the unusual features  has been interpreted within several  different approaches: soft anharmonic nature of the potentials \cite{Klinger_1983}, evolution from a van Hove singularity \cite{schirmacher_1998}, crossover from a minima-dominated phase to a saddle-point-dominated phase \cite{grigera_2003}, proximity of a ``jamming" singularity \cite{Wyart}, or existence of elastic heterogeneities
\cite{Mermet,schirmacher_2006,schirmacher_2007}. The concept of  of elastic heterogeneity in amorphous materials  has been well established thanks to a series of simulation studies \cite{yoshimoto_2004,tsamados_2009,makke_2011}.
Contrary to crystals, amorphous materials exhibit spatially inhomogeneous elastic moduli, with coexistence of hard and soft regions on the scale of a few tenths of particles. Soft regions exhibit even negative elastic moduli \cite{yoshimoto_2004}, which are stabilized by surrounding stable regions with positive moduli.
It is natural that such modulus heterogeneities induce anomalies in vibrational states and thermal properties, and several works have attempted to clarify this connection.
Leonforte \textit{et al.} \cite{leonforte_2005} studied the spatial structure of non-affine displacement fields under deformation, which reflects the elastic heterogeneity, and  found a correlation between that structure and the Boson peak frequency.
Schirmacher \textit{et al.} \cite{schirmacher_2006,schirmacher_2007} proposed a theoretical model, where random spatial fluctuation of shear modulus determines the Boson peak and even the anomalous temperature dependence of thermal conductivity.
A more recent study \cite{marruzzo_2013} showed qualitative agreement between this model and simulation results. 

The aim of this work is to investigate in a more systematic manner this connection, by studying a system in which the heterogeneous nature of the elastic response can be varied continuously by tuning a control parameter.
This is achieved here by progressively introducing a size disorder in a crystalline system, in such a way that the system eventually undergoes structural amorphisation, as shown in Ref. \cite{Bocquet_1992}.
The elastic heterogeneity, vibrational states, and thermal conductivity are monitored simultaneously across this evolution towards the amorphous state.

%%%%%%%%%%%%%%%%%%%%%%%%%%%%%%%%%%%%%%%%%%%%%%%%%%%%%%%%%%%%%%%%%%%%%%%%%%%%%%%%%%%%%%%%%%%%%%%%%%%%%%%%%%%%%%%%%%
\section{System description}
We start from an one-component, perfect face-centered cubic (FCC) crystal in 3-dimensions.
Particles  $i$ and $j$ interact via a soft-core potential $\phi^{ij}=\epsilon(\sigma^{ij}/r^{ij})^{12}$ with $\sigma^{ij} = (\sigma^i +\sigma^j)/2$, where $\sigma^i$ and $\sigma^j$ are their diameters, and $r^{ij}$ is the distance between them.
The potential is cut off and shifted to zero at $r^{ij}=2.5 \sigma^{ij}$. In the initially monodisperse system, $\sigma^i=\sigma$ for all particles.
Owing to the scaling property of the inverse power potential \cite{bernu_1987}, the thermodynamic state depends on a single parameter $\Gamma = \hat{\rho} (\epsilon/k_B T)^{1/4} \sigma^3$, where $k_B$ is Boltzmann constant, $T$ is the temperature, and $\hat{\rho}=N/V$ is the number density.
In the following, $\sigma$, $\epsilon/k_B$, and $\tau=(m\sigma^2/\epsilon)^{1/2}$ (where $m$ is the particle mass) are used as units of length, temperature, and time, respectively.
We have set $\hat{\rho}=1.015$ and the length of the unit cell of the FCC crystal is $a=1.58$.
Most part of our simulations used a  system of $N=4,000$ particles, of linear dimension $L=10a=15.8$.
For the calculation of the vibrational states, larger  systems with  $L$ ranging from $L=12a$ ($N=6,912$) to $30a$ ($N=108,000$) were also considered.
The FCC crystalline system was equilibrated at  $T=0.01$ ($\Gamma=3.21$) by using a $NVT$ MD simulation with periodic boundary conditions. For the monodisperse system, 
the melting and glass transition temperatures have been reported to be $T_m \simeq 0.6$ ($\Gamma_m \simeq 1.15$) and $T_g \simeq 0.2$ ($\Gamma_g \simeq 1.5$), respectively \cite{bernu_1987}.

Disorder was introduced into the crystalline state as described in Ref. \cite{Bocquet_1992}.
Half of the particles, randomly selected,  are assigned to species $1$, the remaining half constitute component  $2$.
The size $\sigma_1$ of species $1$ is gradually reduced below its initial value $1 [\sigma]$, while at the same time, the size $\sigma_2$ of $2$ is increased.
During the process, an ``effective diameter" $\sigma_{\text{eff}}$ of the binary mixture is held fixed at the initial value $1 [\sigma]$.
In an approximate one-component description, $\sigma_{\text{eff}}$ is given by $\sigma_{\text{eff}}^3 = \sum_{\alpha,\beta=1,2} x_\alpha x_\beta \sigma_{\alpha \beta}^3$ \cite{bernu_1987}, where $x_1=x_2=1/2$ are the fractions of the two components.
$\lambda = \sigma_1/\sigma_2 \le 1$ measures the   degree of disorder, and uniquely determines $\sigma_1$ and $\sigma_2$  using the condition $\sigma_{\text{eff}}^3 \equiv 1$.
The coupling parameter $\Gamma$ is replaced  by $\Gamma = \hat{\rho} (\epsilon/k_B T)^{1/4} \sigma^3_\text{eff}$ for the binary mixture, and held fixed  at the initial large value $\Gamma \equiv 3.21$.
In the following calculations, $\lambda$  was decreased starting from the ideal crystal ($\lambda=1$)  by a succession of small steps, $\Delta \lambda = 10^{-4}$.
After each change $\Delta \lambda$, the system was re-equilibrated  using a $NVT$ MD simulation.

%%%%% Figure 1 %%%%%
\begin{figure}
\begin{center}
\includegraphics[scale=1]{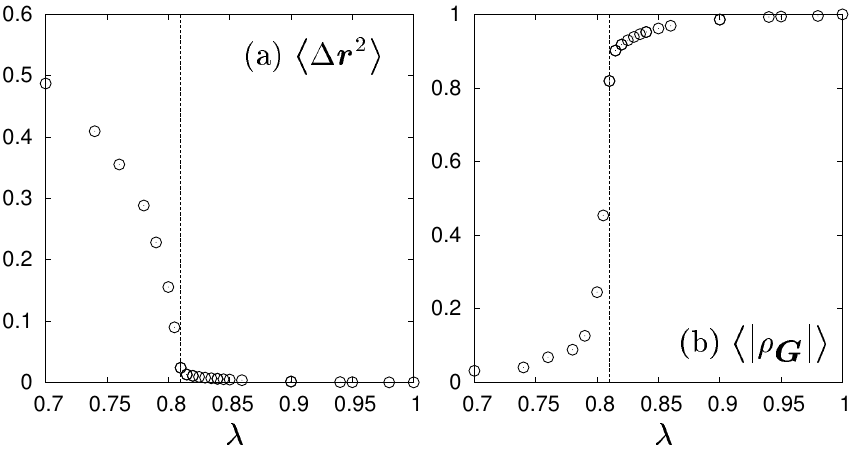}
\end{center}
\vspace*{-6mm}
\caption{
Order parameters across the amorphisation transition.  (a) $\left< \Delta \vec{r}^2\right>$ and (b) $\left< \left| \rho_{\vec{G}} \right| \right>$, versus $\lambda$.
The system undergoes a transition from crystal to amorphous state at $\lambda = \lambda^\ast \simeq 0.81$, indicated by vertical lines.
} 
\label{fig1}
\end{figure}
%%%%%%%%%%%%%%%%%%%%

Upon decreasing $\lambda$, the system undergoes a discontinuous transition toward an amorphous state, as observed (in 2-dimensions) in Ref.~\cite{Bocquet_1992}\footnote{Note that this cannot be considered as a genuine first order phase transition. See Ref.~\cite{Bocquet_1992} for details.}.
The transition point $\lambda^\ast$ is located by monitoring two order parameters, $\left< \Delta r^2 \right> = (1/N) \sum_{j=1}^N( \left< \vec{r}^j \right> - \vec{r}^{j}_{0})^2$ and $\left< \left| \rho_{\vec{G}} \right| \right>=\left< \left| (1/N) \sum_{j=1}^N \exp(i \vec{G}\cdot \vec{r}^j ) \right| \right>$. Here, $\left< \right>$ denotes a time average, $\vec{r}^j$ is the instantaneous position of $j$th particle, $\vec{r}^{j}_{0}$ is the FCC lattice site, and $\vec{G}$ is the smallest reciprocal lattice vector, $\vec{G}=(2\pi/a,2\pi/a,-2\pi/a)$.
Fig.  \ref{fig1} shows $\left< \Delta r^2 \right>$ and $\left< \left| \rho_{\vec{G}} \right| \right>$ versus $\lambda$.
The transition takes place  at $\lambda \simeq 0.81$, where both $\left< \Delta r^2 \right>$ and $\left< \left| \rho_{\vec{G}} \right| \right>$ exhibit  discontinuous jumps.
We confirmed that  $\lambda^\ast$ does not depend on the initial repartition  of the two species on the lattice.
For $1 > \lambda \ge \lambda^\ast \simeq 0.81$, the system is in  a chemically disordered crystalline state, with non-zero Bragg peaks for  the number  density. For $\lambda^\ast > \lambda \ge 0.7$, the system is in an amorphous state, with complete loss of periodicity.

%%%%% Figure 2 %%%%%
\begin{figure*}
\begin{center}
\includegraphics[scale=0.90]{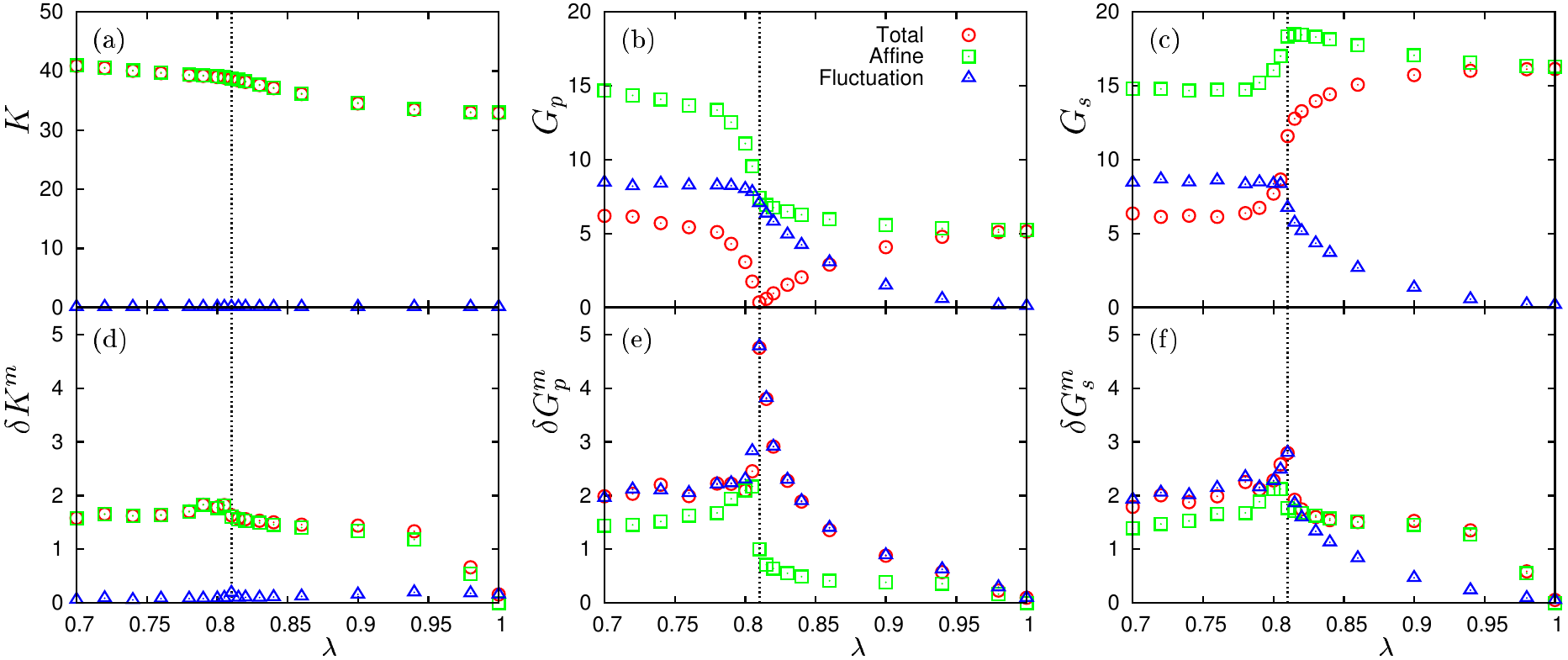}
\end{center}
\vspace*{-6mm}
\caption{
The macroscopic values (average values), (a) $K$, (b) $G_p$, and (c) $G_s$, and the standard deviations, (d) $\delta K^m$, (e) $\delta G_p^m$, and (f) $\delta G_s^m$, versus the disorder parameter  $\lambda$.
Circles, squares, and triangles represent values of the total modulus $C^m$, the affine term $C^m_A$, and the fluctuation term $C^m_F$, respectively.
The vertical lines indicate the transition point $\lambda=\lambda^\ast \simeq 0.81$.
} 
\label{fig2}
\end{figure*}
%%%%%%%%%%%%%%%%%%%%

\section{Elastic heterogeneity}
Inhomogeneous systems like glasses or complex crystals exhibit heterogeneous, scale-dependent distributions of local elastic moduli, which can be determined  using several methods \cite{yoshimoto_2004,tsamados_2009,makke_2011, Mizuno_2013}.
Here,  the system was divided into $20 \times 20 \times 20$  cubic domains with a linear size $W=2a=3.16$, 
and the local elastic moduli were obtained from equilibrium fluctuation formulae \cite{lutsko_1988}. (Details about the calculations are included in Supplemental Material \cite{supplement}.)
For each subvolume,  one local bulk modulus and five local shear moduli  are defined, corresponding to the response to isotropic bulk compression and volume-conserving shear deformations, respectively \cite{Mizuno_2013}.
Of the  five shear moduli, two correspond to ``pure" shear deformations (plane and triaxial strain deformations), and the remaining three are related to ``simple" shear deformations.
In the following, the local bulk, pure shear, and simple shear moduli are denoted by  $K^m$, $G_p^m$, and $G_s^m$, respectively (where ``m"  indicates a  local quantity).

For each local modulus $C^m=K^m, G_p^m, G_s^m$,  the probability distribution function $P(C^m)$, which turns out to be closely Gaussian, was obtained by sampling the different cubic subvolumes. Note that $P(G_p^m)$ and $P(G_s^m)$ are different in cubic crystals, while they coincide with each other in isotropic glasses. 
 Each modulus is the sum of four contributions, the Born term ${C}^m_B$,  the kinetic contribution ${C}^m_K$,  the pressure correction term ${C}^m_C$, and the fluctuation term $-{C}^m_F$ (See Supplemental Material \cite{supplement}). The sum of the first three terms, denoted by ${C}^m_A$  in Fig. \ref{fig2}, corresponds to the response of a system which deforms affinely at all scales; The fluctuation term $-{C}_F^m$ is a negative correction that accounts for the non-affinity of the deformation at small scales, and in disordered systems  becomes comparable in magnitude to  $C_A^m$ \cite{tanguy_2002}.
The average value, $C=\int C^m P(C^m) dC^m$, and standard deviation, 
$\delta C^m = \sqrt{\int (C^m-C)^2 P(C^m)dC^m}$, are reported in figure  \ref{fig2}.
The average value, $C=K, G_p, G_s$, coincides with the ``macroscopic" modulus, denoted without ``m".
In the initial crystalline state $\lambda=1$, $K > G_s > G_p$, while $K > G_p \simeq G_s$ for the isotropic amorphous states $\lambda \le 0.78$. For a system with inverse power interactions, the bulk modulus $K$ is determined by the affine $K_A$ only, i.e., $K = K_A$ ($K_F =0$) (see Fig. \ref{fig2}(a)).
On the other hand, the non-affine components, $G_{pF}$ and $G_{sF}$, get  progressively larger with decreasing $\lambda$.
Remarkably, around the transition $\lambda^\ast$, $G_{pF}$ reaches the affine $G_{pA}$, so that  the total $G_p$ vanishes, $G_p \simeq 0$.
Therefore, the transition can be described as an elastic instability associated with the shear modulus $G_p$.
Below  the transition ($\lambda < \lambda^\ast$),  the material becomes rapidly isotropic, as it is manifested by the convergence  $G_p \simeq G_s$.

Across the amorphisation transition, the elastic heterogeneity, characterized by the standard deviation of the moduli, undergoes important changes, starting from a spatially homogeneous modulus distribution at $\lambda=1$.
As $\lambda$ decreases from $1$ to $0.9$, $\delta K^m$ and $\delta G^m_s$ increase monotonically, mainly in relation to an increase in $\delta K^m_A$ and $\delta G^m_{sA}$, respectively. In the same range, $\delta G^m_p$ is already dominated by the heterogeneity in the fluctuation term $\delta G^m_{pF}$.
As $\lambda$ approaches $\lambda^\ast$, this heterogeneity increases rapidly, and the spatial distribution of $G^m_p$ becomes extremely  heterogeneous with a nearly zero average value, $G_p \simeq 0$ and a large standard deviation, $\delta G^m_p \simeq 5$.
 Below  the transition point $\lambda^\ast$, $\delta G_p^m$ and $\delta G_s^m$ immediately converge to a very similar value, $\delta G_{p,s}^m \simeq 2$. 
In the following,  these changes in elastic properties will be correlated with the evolution of the vibrational and thermal properties. 

%%%%%%%%%%%%%%%%%%%%%%%%%%%%%%%%%%%%%%%%%%%%%%%%%%%%%%%%%%%%%%%%%%%%%%%%%%%%%%%%%%%%%%%%%%%%%%%%%%%%%%%%%%%%%%%%%%

%%%%% Figure 3 %%%%%
\begin{figure}
\begin{center}
\includegraphics[scale=1.25]{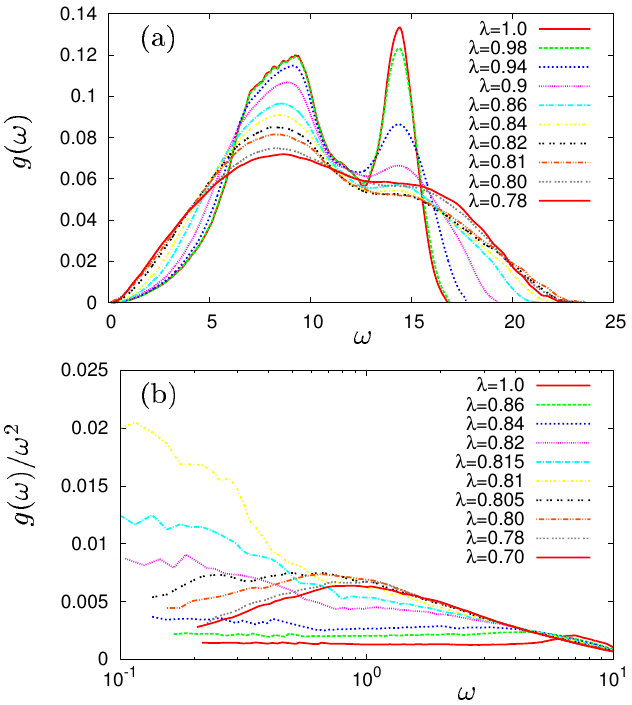}
\end{center}
\vspace*{-6mm}
\caption{
(a) Density of states $g(\omega)$ and (b) reduced density of states $g(\omega)/\omega^2$ for various $\lambda$s.
} 
\label{fig3}
\end{figure}
%%%%%%%%%%%%%%%%%%%%%

\section{Vibrational states}
A standard normal modes analysis was performed by diagonalizing the Hessian matrix for all considered values of the disorder parameter, in the range $1\ge \lambda \ge 0.7$. The resulting vibrational densities of states (VDOS)   $g(\omega)$ are shown in figure \ref{fig3}(a).
In the crystalline state $\lambda=1$, a ``transverse branch" in the range $\omega\in [7:9.5]$ and a ``longitudinal branch" around $\omega = 14.5$ can be clearly identified in the VDOS.
As $\lambda$ decreases, these contributions from well identified phonon branches tend to loose their identity.
In particular, the peak associated with the longitudinal branch is already suppressed at $\lambda = 0.9$, which is correlated with the appearance of significant heterogeneities of the bulk modulus $K^m$ and of the  shear modulus $G^m_s$.
These heterogeneities induce a broadening of the high $\omega$ modes, which leads to the reduction of the intensity of the longitudinal peak.

%%%%%% Figure 4 %%%%%
\begin{figure}
\begin{center}
\includegraphics[scale=1.09]{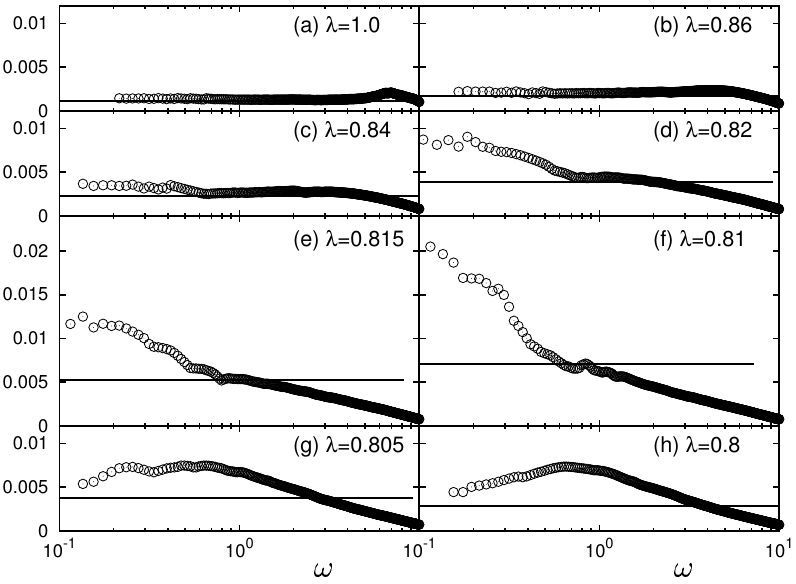}
\end{center}
\vspace*{-6mm}
\caption{
Reduced density of state $g(\omega)/\omega^2$ compared to the Debye-model prediction $g_D(\omega)/\omega^2 = 3/\omega_D^3$.
Symbols and lines represent $g(\omega)/\omega^2$ and  $g_D(\omega)/\omega^2$, respectively.
} 
\label{fig4}
\end{figure}
%%%%%%%%%%%%%%%%%%%%

The heterogeneity in low shear modulus $G_p^m$, on the other hand, is well correlated with the behaviour of the VDOS in the  low $\omega$ region. This correlation is most easily seen by 
considering  the reduced VDOS,  $g(\omega)/\omega^2$, shown in Figs. \ref{fig3}(b) and \ref{fig4}.
In Fig \ref{fig4}, $g(\omega)/\omega^2$ is compared to the Debye-model prediction, $g_D(\omega)/\omega^2 = 3/\omega_D^3$, where the Debye frequency $\omega_D$ is calculated from the macroscopic moduli, $K$, $G_p$, and $G_s$ \cite{Jasiukiewicz_2003}.
At $1 \ge \lambda \ge 0.84$, $g(\omega)/\omega^2$ coincides with $g_D(\omega)/\omega^2$, indicating that low $\omega$ excitations are not affected by the modulus heterogeneities.
An excess of modes in $g(\omega)/\omega^2$ over $g_D(\omega)/\omega^2$ starts to appear around $\lambda =0.82$. This excess is particularly pronounced at very low frequencies, and increases with the spatial heterogeneity of $G_p^m$ as $\lambda$ approaches $\lambda^\ast$.
Below the transition $\lambda^\ast$, the excess of modes drops immediately, accompanying a rapid decrease in the heterogeneity of   $G_p^m$.
At $\lambda \le 0.8$, a  typical Boson peak, with an amplitude comparable to what is typically observed in soft sphere or Lennard-Jones~\cite{monaco_2009} glasses, is formed around $\omega = \omega_\text{BP} \simeq 1$.
Note that the excess of modes in the disordered crystals close to $\lambda^\ast$ is observed at much lower frequencies, presumably down to zero frequency at the transition point, in relation to the vanishing of $G_p$ at the transition.
Altogether, the position and amplitude of  the peak appear to be controlled by the shear modulus average value and heterogeneity, in the disordered crystal as well as in the amorphous state. This is qualitatively consistent with the theoretical ideas of Refs.~\cite{schirmacher_2006,schirmacher_2007}, and a more quantitative analysis will be presented in future work. 

%%%%% Figure 5 %%%%%
\begin{figure}
\begin{center}
\includegraphics[scale=1.13]{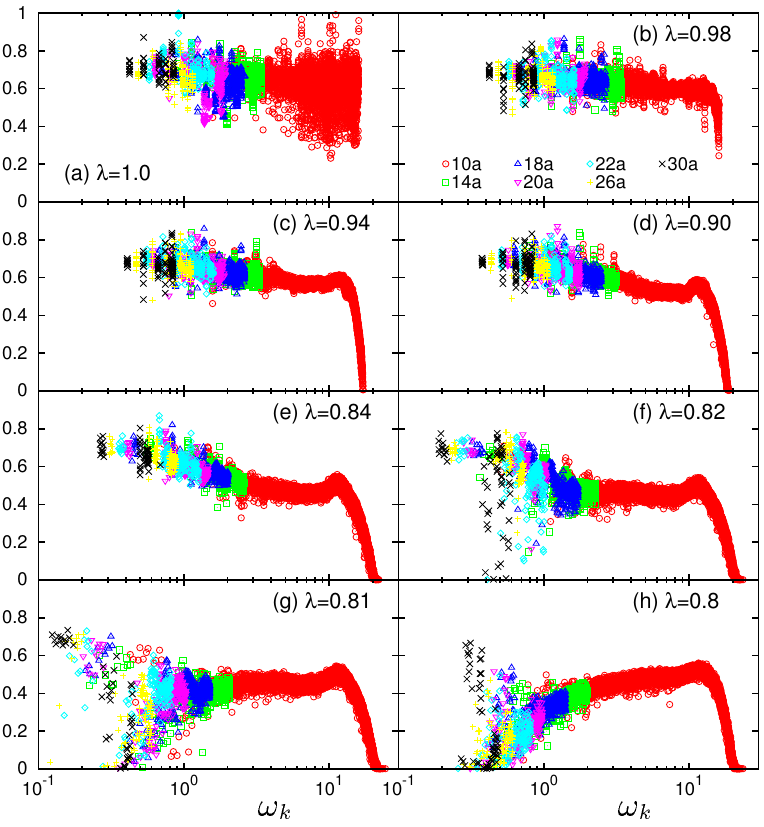}
\end{center}
\vspace*{-8mm}
\caption{
Participation ratio $PR_k$ versus $\omega_k$.
To access lower $\omega_k$ modes,  several systems with sizes $L$ ranging from $L=10a$ ($N=4,000$) to $30a$ ($N=108,000$) have been considered.} 
\label{fig5}
\end{figure}
%%%%%%%%%%%%%%%%%%%%

The participation ratio, $PR_k = (1/N) \left[ \sum_{j=1}^N ( \vec{e}^j_k \cdot \vec{e}^j_k )^2 \right]^{-1}$, where $k$ is the eigen-mode number, and $\vec{e}^j_k$ is the polarization vector, is shown with the eigen-frequency $\omega_k$ in Fig. \ref{fig5}. (See Supplemental Material \cite{supplement} for the eigen-vectors spatial structure.)
As extreme cases, $PR_k=2/3$ for an ideal standing plane wave and $PR_k \simeq 1/N$ for an ideal mode involving only one particle \cite{mazzacurati_1996}.
As $\lambda$ decreases from $1$ to $0.9$, high $\omega$ modes become localized with low $PR_k$ values. This corresponds to the disappearance of the longitudinal peak in the VDOS, and can be related to the heterogeneities of $K^m$ and $G_s^m$.
On the other hand, low $\omega$  quasi-localized modes  \cite{shober_2004} start to become apparent at $\lambda \le 0.82$.
At $\lambda=0.8$, i.e., in the amorphous state, the result is consistent with that reported for a Lennard-Jones glass \cite{mazzacurati_1996}.
Clearly, the appearance of these quasi-localized modes is correlated with that of an excess VDOS, and to the sharp increase in the heterogeneity of $G_p^m$.
This confirms the recent findings in simulation \cite{shintani_2008} and experiment \cite{tan_2012} which demonstrated  the transverse nature of low frequency excess modes, which were assigned to the coupling of transverse vibrational modes with defective soft structures.

%%%%%%%%%%%%%%%%%%%%%%%%%%%%%%%%%%%%%%%%%%%%%%%%%%%%%%%%%%%%%%%%%%%%%%%%%%%%%%%%%%%%%%%%%%%%%%%%%%%%%%%%%%%%%%%%%%

\section{Thermal conductivity}
\begin{figure}
\begin{center}
\includegraphics[scale=1.25]{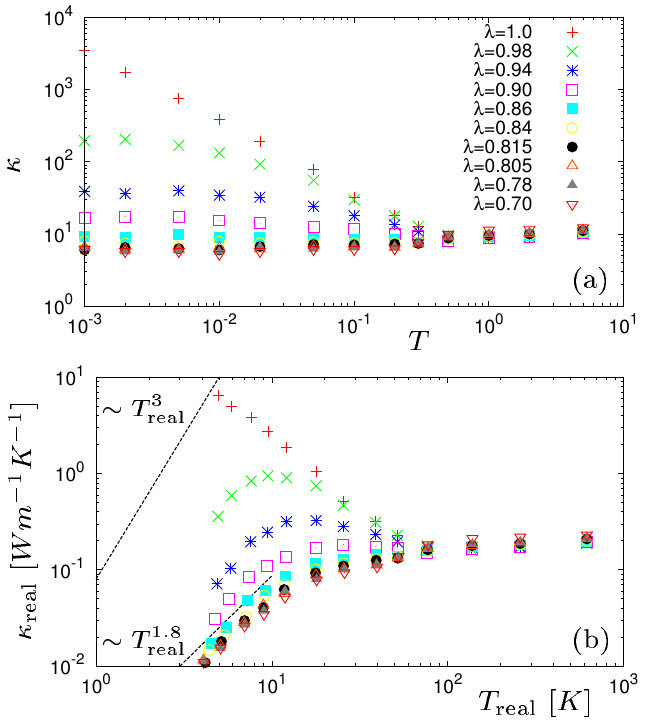}
\end{center}
\vspace*{-8mm}
\caption{
The temperature dependence of thermal conductivity $\kappa$ for $\lambda=1$ to $0.7$.
(a) The values obtained from MD simulations and (b) the values with quantum corrections.
In (b), $T_\text{real}$ and $\kappa_\text{real}$ are measured in units of $K$ and $Wm^{-1}K^{-1}$, respectively.
} 
\label{fig6}
\end{figure}

The thermal conductivity $\kappa$ was obtained from 
 $NVE$ MD simulations, using the Green-Kubo formulation \cite{macgaughey_2006}: $\kappa = ({1}/{3k_B T^2 V}) \int_{0}^\infty {\left< \vec{J}(t) \cdot \vec{J}(0) \right>} dt$, where $\vec{J}$ is the heat current vector.
Figure \ref{fig6}(a) shows the  temperature dependence of $\kappa$ at the indicated values of  $\lambda$.
In the crystalline state $\lambda=1$, $\kappa$ decreases with  increasing $T$, due to the increase of  anharmonicity.
As $\lambda$ decreases, the disorder scatters heat carriers and reduces $\kappa$ steeply.
The reduction of $\kappa$ saturates around $\lambda =0.86$, i.e., \textit{before} a significant excess of modes is observed in the VDOS. 
For  $\lambda \le 0.86$, $\kappa$ is insensitive to $T$, which means that  disorder dominates  anharmonic effects.

%%%%% Figure 6 %%%%%
%%%%%%%%%%%%%%%%%%%%

The above behavior of $\kappa$ can be rationalized  by using the simple formula: $\kappa = (1/3) C v \ell$ \cite{macgaughey_2006}, where $C$ is the heat capacity per unit volume, $v$ and $\ell$ are the ``average" velocity and mean free path of the heat carriers, respectively.
$C$ and $v$ are taken as constants, and the attenuation rate $\sim 1/\ell$, a function of $T$ and $\lambda$, is decomposed into two components, anharmonic $1/\ell_{\text{anh}}$ and disorder $1/\ell_{\text{dis}}$, which respectively depend on $T$ and $\lambda$: $1/\ell(T,\lambda) = 1/\ell_{\text{anh}}(T) + 1/\ell_{\text{dis}}(\lambda)$ \cite{masciovecchio_2006}.
 $1/\ell_{\text{dis}}$ is expected to be increased  by the modulus heterogeneities.
In the perfect crystal ($\lambda=1$), $1/\ell_{\text{anh}} \gg 1/\ell_{\text{dis}}$, leading to $\ell \simeq \ell_{\text{anh}}$.
In this case, the anharmonic effect reduces $\ell$ with increasing $T$ and causes the reduction of $\kappa$.
As $\lambda$ decreases, $1/\ell_{\text{dis}}$ becomes large, and $1/\ell_{\text{anh}} \ll 1/\ell_{\text{dis}}$ at $\lambda \le 0.86$.
Comparing with the results reported in Fig. \ref{fig2}, it appears that the  increase of $1/\ell_{\text{dis}}$ is mainly correlated to  the heterogeneities of $K^m$ and $G_s^m$.
At $\lambda =0.86$, we measured $\ell =3\kappa/Cv \sim 1[\sigma]$, the order of particle diameter, i.e., $\ell$ and $\kappa$ reach a ``minimum" value, insensitive to temperature.
At this stage, even if  further  disorder  makes local moduli more heterogeneous, $\kappa$ does not decrease any more.
Actually, in such a strong scattering state with an extremely short $\ell$, heat carriers are no longer propagating phonons, but non-propagating or quasi-stationary excitations, sometimes called diffusons \cite{allen_1993}.

In  the low temperature regime, the calculations presented above are based on a classical approach, where the heat capacity $C$ is independent of $T$.
However, in a real system, $C$ depends on $T$, e.g., $C \sim T^3$ for the Debye model, due to quantum effects.
In order to assess qualitatively how the results would be modified by quantum effects, a simple quantum correction, often used in previous works \cite{macgaughey_2006}, was applied. (See Supplemental Material \cite{supplement} for details.)
The correction maps ``classical" values of $T_{\text{MD}}$ and $\kappa_{\text{MD}}$, obtained from MD simulations, onto ``real" values of $T_{\text{real}}$ and $\kappa_{\text{real}}$.
The mapping of $\kappa_{\text{MD}}$ onto $\kappa_{\text{real}}$ is $\kappa_{\text{real}} = ({C_{\text{real}}}/{C_{\text{MD}}}) \kappa_{\text{MD}}$, which takes into account the ``real" $T$ dependence of the heat capacity $C_{\text{real}}$.
Note that $C_\text{real}$ can be calculated from the density of state $g(\omega)$.
Temperature mapping can be performed in two ways, with or without inclusion of the ``zero point energy".
In the present study, the  correction without inclusion of the zero point energy was applied \cite{macgaughey_2006}.

Figure \ref{fig6}(b) shows the $T_\text{real}$ dependence of $\kappa_\text{real}$.
Here, the units of argon, $\sigma =3.405 \mathrm{\AA}$, $\epsilon/k_B = 125.2 K$, and $\tau = 2.11 \text{ps}$, were used.
The values of the crystalline state $\lambda=1$ are comparable with experimental values for solid argon \cite{christensen_1975}.
For $\lambda \le 0.90$,  the $T$ dependence becomes similar to the one observed in amorphous materials, with a power-law like increase $\kappa_\text{real} \sim T^{1.8}$ at low temperatures ($T<10 [K]$), followed by a  plateau above $T \sim 10 [K]$.
Note that $\kappa$ displays this  ``glassy'' behaviour already in the disordered crystalline states with $0.90 \ge \lambda \ge \lambda^\ast \simeq 0.81$.
This observation is consistent with experimental results reported  by Cahill \textit{et al.} \cite{cahill_1992}.
These authors  measured the thermal conductivity of disordered crystals, for example $(\text{KBr})_{1-x} (\text{KCN})_x$, by changing the composition $x$.
$x=0$ corresponds to the crystalline state, and as $x$ increases to $0.5$, more disorder is introduced, so that $x$ plays a role similar to $\lambda$. As $x$ was increased, $\kappa$ was found to decrease  steeply and  to exhibit a glass-like $T$ dependence even in disordered crystals.

%%%%%%%%%%%%%%%%%%%%%%%%%%%%%%%%%%%%%%%%%%%%%%%%%%%%%%%%%%%%%%%%%%%%%%%%%%%%%%%%%%%%%%%%%%%%%%%%%%%%%%%%%%%%%%%%%%
\section{Conclusions and remarks}
In this work, we have used the flexibility offered by a numerical approach to study the correlations between elastic heterogeneity, vibrational states, and thermal conductivity in a model in which those quantities can be controlled continuously.
We have been able to investigate in a unified framework the cases of the perfect crystal state, increasingly disordered crystals, and the isotropic amorphous solid state.
We find that the heterogeneities of the bulk modulus $K^m$ and the highest shear modulus $G^m_s$ are well correlated with change in nature of the high frequency vibrational states.
As $\lambda$ decreases from $1$ to $0.9$, $K^m$ and $G^m_s$ become heterogeneous progressively, while the high $\omega$ modes loose their propagative character, leading to the reduction of the longitudinal branch in $g(\omega)$.
The appearance of these heterogeneities is also correlated to a steep decrease of thermal conductivity $\kappa$, as the high frequency modes become non-propagating.

On the other hand, the lowest shear modulus $G^m_p$, which vanishes at the amorphisation point, was observed to be correlated with the low $\omega$ vibrational states.
As $\lambda$ goes to $\lambda^\ast$, $G^m_p$ gets increasingly  heterogeneous.
Accompanying this heterogeneity, the excess of modes in $g(\omega)/\omega^2$ becomes pronounced, and at the same time, the low $\omega$ modes get localized.
After the transition $\lambda^\ast$, the heterogeneity of $G_p^m$ converges to that of amorphous states immediately, and the excess peak in $g(\omega)/\omega^2$ also converges to the usual Boson peak. 

We finalize this letter with a few remarks.
The diverging behavior of $g(\omega)/\omega^2$ as $\lambda \rightarrow \lambda^\ast$ is very similar as that reported in jamming systems \cite{silbert_2005}, where as the system goes to an unjamming point, $g(\omega)$ approaches a non-zero value, i.e., $g(\omega)/\omega^2$ diverges, in the limit of zero $\omega$.
However, in our system, $g(\omega)$ always approaches a zero value even at $\lambda =0.81 \simeq \lambda^\ast$, where $g(\omega) \sim \omega^{1.5}$ is observed in our $\omega$ window.
As one possibility, the difference may come from that only $G_p^m$ is heterogeneous around a vanishing value, $G_p \simeq 0$, in our system, whereas in jamming systems, both $G_p^m$ and $G_s^m$ vanish simultaneously.

A most striking result is that $\kappa$ is almost unchanged through the transition $\lambda^\ast$, indicating that $\kappa$,  in a classical system,  is not directly  related to excess modes and low $\omega$ localizations.
This decoupling may be partially a result of the classical nature of our simulations, where all vibrational modes are excited equally \cite{turney_2009}.
In real quantum systems, only low $\omega$ modes are excited according to the Bose-Einstein statistics, so that $\kappa$  is expected to be more influenced by the low $\omega$ excitations.
In fact, calculations using a quantum approach based on the mode diffusivity \cite{allen_1993} suggested a correlation between the Boson peak and the anomalous thermal conductivity \cite{xu_2009}.
This correlation is not apparent here, and from the classical calculation it appears that all the modes (including the low frequency ones) have already lost their heat transport capability before they become localized and the amorphisation threshold is reached. 

As a final remark, we emphasize that  the modulus distribution, the vibrational states, and the  thermal conductivity are almost insensitive to disorder once the amorphisation threshold has been reached. 
In contrast, we observed important quantitative changes in the  disordered crystalline states $1 > \lambda > \lambda^\ast$.
This  suggests that introducing  disorder into crystals may be a valuable  strategy to control material properties, rather than attempting to control the amorphous state. In fact, synthetic nanostructuring in crystals is considered to be most promising way for controlling thermal conductivity \cite{ma_2013}.

%%%%%%%%%%%%%%%%%%%%%%%%%%%%%%%%%%%%%%%%%%%%%%%%%%%%%%%%%%%%%%%%%%%%%%%%%%%%%%%%
\vspace{1.0cm}

\section{Supplemental Material}

This is supplemental material which integrates the published paper.
\vspace{0.5cm}

\noindent
{\bf Local elastic modulus.-} The local elastic modulus ${C}^m_{ijkl}$ is defined as the first-derivative of the local stress ${\sigma}^m_{ij}$ with respect to the local strain ${\epsilon}^m_{kl}$ \cite{Mizuno_2013}. One possible implementation of the direct calculation of ${C}^m_{ijkl}$ is in terms of the equilibrium fluctuation formulae~\cite{Mizuno_2013,lutsko_1988}. In Ref.~\cite{Mizuno_2013}, we referred to this method as the ``Fully local approach", which proceeds as follows.

The simulation box is partitioned into $20^3$ cubic domains, identified by the index ``$m$" and of linear size $W=2a=3.16$. Here, $a=1.58$ is the length of the unit cell for the FCC crystal. For each cube $m$, the local modulus tensor $C_{ijkl}^m$ is obtained as the sum of four terms: the Born term, $C_{Bijkl}^{m}$, the kinetic contribution, $C_{Kijkl}^{m}$, the pressure correction, $C_{Cijkl}^{m}$, and the fluctuation term, $-C_{Fijkl}^{m}$ :
\begin{equation}
C_{ijkl}^m  = C_{Bijkl}^{m} + C_{Kijkl}^{m} + C_{Cijkl}^{m} - C_{Fijkl}^{m}
 = C_{Aijkl}^{m} - C_{Fijkl}^{m}.
\label{equation1}
\end{equation}
Here, $C_{Aijkl}^{m}=C_{Bijkl}^{m} + C_{Kijkl}^{m} + C_{Cijkl}^{m}$ corresponds to the response of a system which deforms affinely, while $-C_{Fijkl}^{m}$ accounts for the non-affinity of the deformation. The terms in Eq. (\ref{equation1}) are evaluated via the equations:
\begin{equation}
\begin{aligned}
C_{Bijkl}^{m} & = \frac{1}{W^3} \Biggl< \sum_{a<b} \left( \derri{\phi^{ab}}{{r^{ab}}^2} - \frac{1}{r^{ab}}\deri{\phi^{ab}}{r^{ab}} \right) \\
& \qquad \qquad \qquad \times \frac{r^{ab}_i r^{ab}_j r^{ab}_k r^{ab}_l}{ {r^{ab}}^2 }\frac{q^{ab}}{r^{ab}} \Biggr>, \\
C_{Kijkl}^{m} & = 2\left< \hat{\rho}^m \right>T (\delta_{ik} \delta_{jl} + \delta_{il}\delta_{jk}), \\
C_{Cijkl}^{m} & = -\frac{1}{2} \big( 2\left<\sigma^m_{ij}\right>\delta_{kl}-\left<\sigma^m_{ik}\right>\delta_{jl}-\left<\sigma^m_{il}\right>\delta_{jk} \\
& \qquad \quad - \left<\sigma^m_{jk}\right>\delta_{il} -\left<\sigma^m_{jl}\right>\delta_{ik} \big), \\
C_{Fijkl}^{m} & = \frac{V}{T} [\left< \sigma^m_{ij} \sigma_{kl} \right>-\left< \sigma^m_{ij} \right>\left< \sigma_{kl} \right>].
\end{aligned}
\end{equation}
Here, $\left< \right>$ represents the ensemble average, $\hat{\rho}^m$ is the local number density in the cube $m$, $r^{ab}_i$ denotes the vector joining particles $a$ and $b$, and $r^{ab}$ is their distance.
The quantity $q^{ab}$ represents the length of the line segment $r^{ab}_i$ which is located inside the cube $m$. If the vector $r^{ab}_i$ is located outside cube $m$, it follows $q^{ab}=0$. The term $q^{ab}/r^{ab}$ therefore determines the contribution of each pairwise interaction to the Born term $C_{Bijkl}^{m}$. The local stress $\sigma_{ij}^m$ and the global stress $\sigma_{ij}$ are calculated as
\begin{equation}
\begin{aligned}
\sigma_{ij}^m &= -\hat{\rho}^m T \delta_{ij} + \frac{1}{W^3} \sum_{a<b} \deri{\phi^{ab}}{r^{ab}} \frac{r^{ab}_i r^{ab}_j}{r^{ab}}\frac{q^{ab}}{r^{ab}}, \\
\sigma_{ij} & = \frac{1}{V} \sum_{m} W^3 \sigma^m_{ij} = -\hat{\rho} T \delta_{ij} + \frac{1}{V} \sum_{a<b} \deri{\phi^{ab}}{r^{ab}} \frac{r^{ab}_i r^{ab}_j}{r^{ab}},
\end{aligned}
\end{equation}
where $\hat{\rho}=1.015$ is the global number density.

%%%%%%%%%%%%%%%%%%%%%%%% Supplementary Figure1 %%%%%%%%%
\begin{figure*}[t]
\begin{center}
\includegraphics[scale=1]{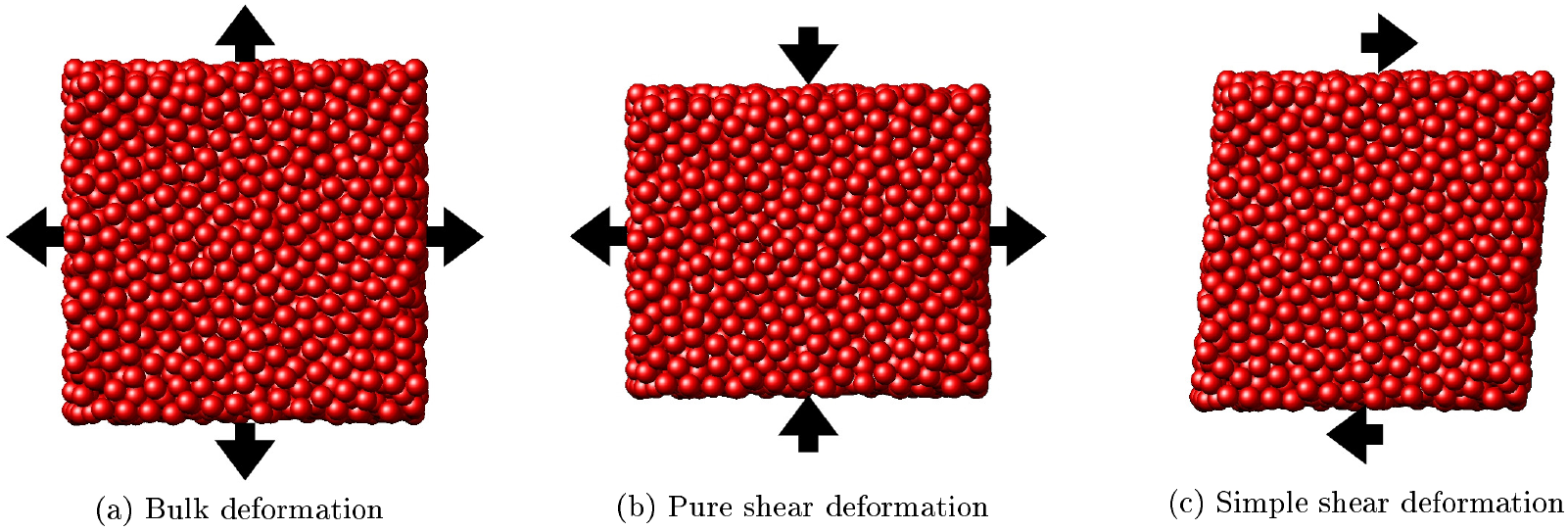}
\end{center}
\vspace*{-7mm}
\caption{Schematic illustration of (a) bulk deformation, (b) pure shear deformation, and (c) simple shear deformation.
}
\label{sfig1}
\end{figure*}
%%%%%%%%%%%%%%%%%%%%%%%%%%%%%%%%%%%%%%%%%%%%%%%%%%%%%%%%

%%%%%%%%%%%%%%%%%%%%%%%% Supplementary Figure2 %%%%%%%%%
\begin{figure*}
\vspace*{1cm}
\begin{center}
\includegraphics[scale=1]{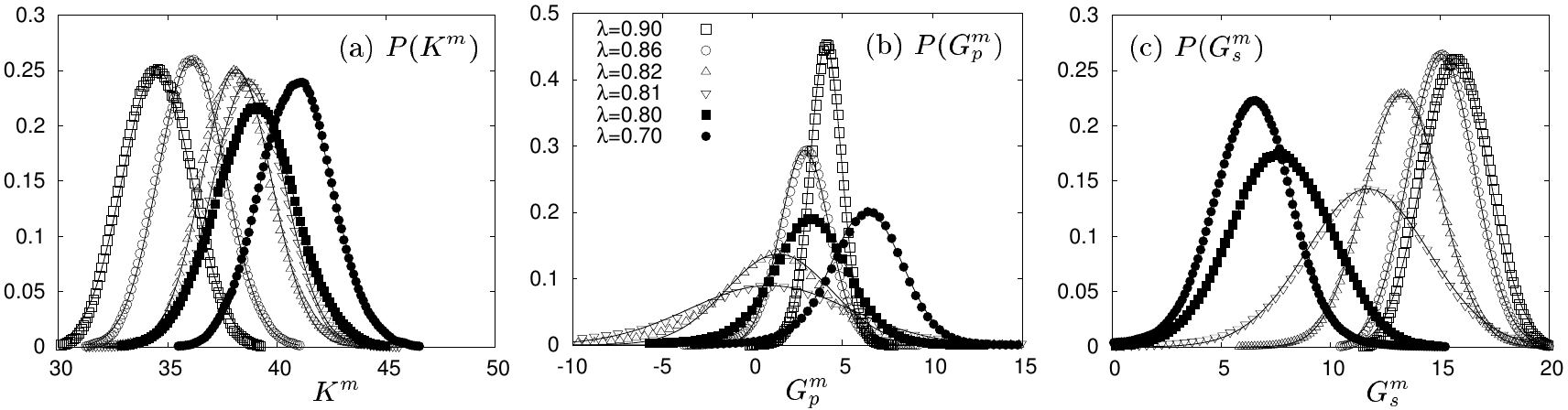}
\end{center}
\vspace*{-7mm}
\caption{Probability distribution of (a) bulk modulus $K^m$, (b) pure shear modulus $G_p^m$, and (c) simple shear modulus $G_s^m$. We also show the Gaussian distribution functions fitted to the calculated probability distributions (solid lines).
}
\label{sfig2}
\end{figure*}
%%%%%%%%%%%%%%%%%%%%%%%%%%%%%%%%%%%%%%%%%%%%%%%%%%%%%%%%

From the modulus tensor $C^m_{ijkl}$, we obtain the local bulk modulus $K^m$ and five local shear moduli $G^m_1$, $G^m_2$, $G^m_3$, $G^m_4$, and $G^m_5$ from the following equations~\cite{Mizuno_2013} :
\begin{equation}
\begin{aligned}
3K^m &= (C^m_{xxxx}+C^m_{yyyy}+C^m_{zzzz}+C^m_{xxyy}+C^m_{yyxx} \\
  & \quad \ +C^m_{xxzz}+C^m_{zzxx}+C^m_{yyzz}+C^m_{zzyy})/3, \\
2G^m_1 & = (C^m_{xxxx}+C^m_{yyyy}-C^m_{xxyy}-C^m_{yyxx})/2, \\
2G^m_2 & = (C^m_{xxxx}+C^m_{yyyy}+4C^m_{zzzz}+C^m_{xxyy}+C^m_{yyxx} \\
    &  \quad -2C^m_{xxzz}-2C^m_{zzxx}-2C^m_{yyzz}-2C^m_{zzyy})/6, \\
2G^m_3 & = C^m_{xyxy}, \\
2G^m_4 & = C^m_{xzxz}, \\
2G^m_5 & = C^m_{yzyz},
\end{aligned} 
\label{equation2}
\end{equation}
Here, $G^m_1$ and $G^m_2$ correspond to ``pure" shear deformations (plane and tri-axial strain deformations), and the remaining $G^m_3$, $G^m_4$, and $G^m_5$ are related to ``simple" shear deformations. We give a schematic illustration for bulk deformation, pure shear deformation, and simple shear deformation in Fig. \ref{sfig1}. In the Letter, we represent $G^m_p$ for $G^m_1$ and $G^m_2$, and $G^m_s$ for $G^m_3$, $G^m_4$, and $G^m_5$.

In Fig. \ref{sfig2}, we show the probability distributions of local elastic moduli $K^m$, $G^m_p$, and $G^m_s$ at the indicated values of $\lambda$. The distributions are well described by Gaussian fitting functions (solid lines). From the distributions, we extract the average values, $K$, $G_p$, and $G_s$, and the standard deviations, $\delta K^m$, $\delta G_p^m$, and $\delta G_s^m$, which are discussed in the Letter.

%%%%%%%%%%%%%%%%%%%%%%%%%%%%%%%%%%%%%%%%%%%%%%%%%%%%%%%%%%%%%%%%%%%%%%%%%%%%%%%%%%%%%%%%%%%%%%%%%%%%%%%%%%%%%%%%%%%%%%%%%
\vspace{1.cm}

\begin{figure*}[p]
\begin{center}
\includegraphics[scale=1]{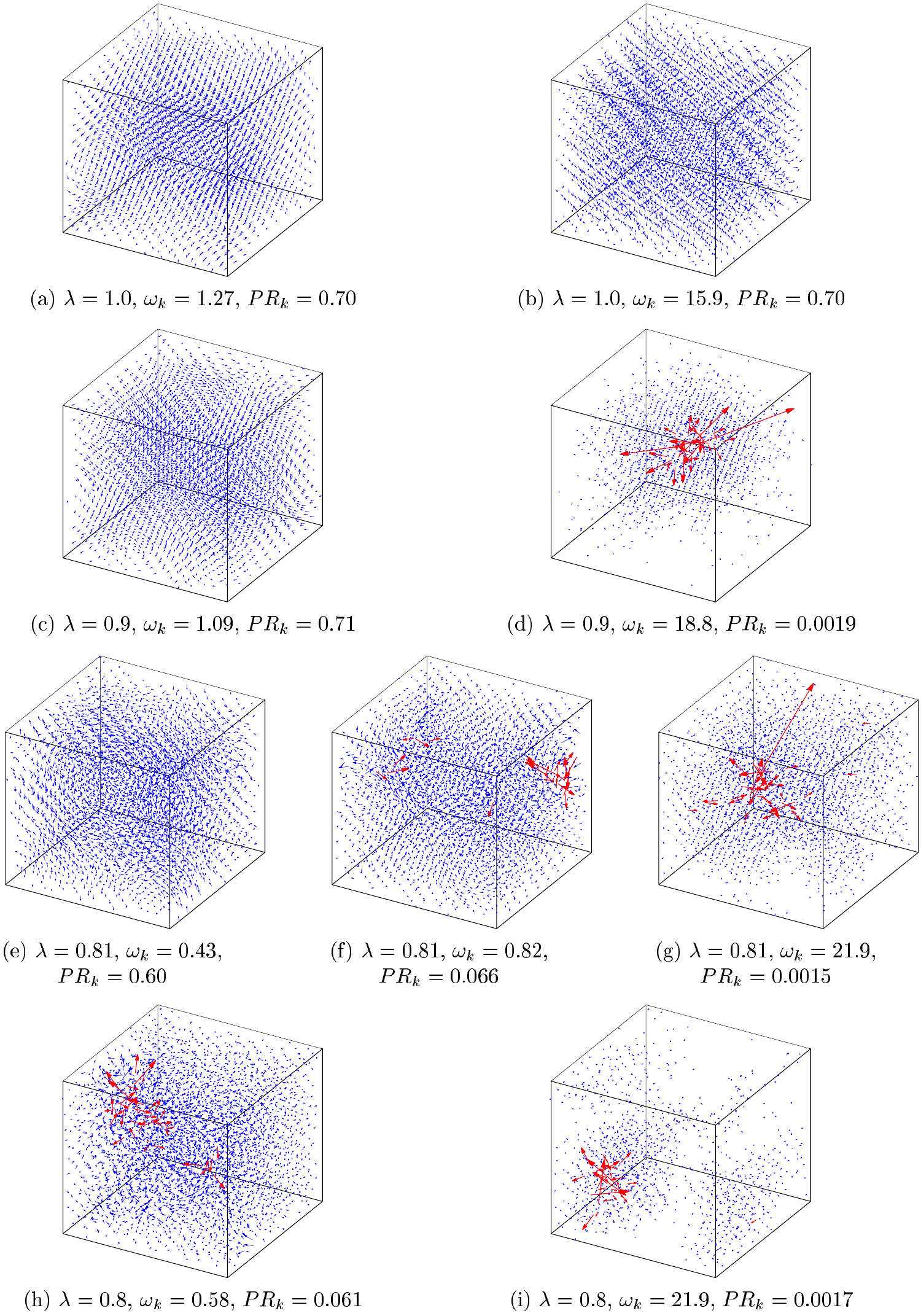}
\end{center}
\vspace*{-5mm}
\caption{Displacement field corresponding to a few eigen-vectors of the Hessian matrix: (a), (b) for $\lambda=1.0$, (c), (d) for $\lambda=0.9$, (e), (f), (g) for $\lambda=0.81$, and (h), (i) for $\lambda=0.8$.
Arrows correspond to the displacements of particles We plotted the eigen-vectors of the particles and are multiplied by a factor of 10, for clarity. We also indicate the eigen-frequency $\omega_k$ and the participation ratio $PR_k$ for each considered mode.
In (d), (f), (g), (h), and (i), corresponding to localized modes with $PR_k < 0.1$, the largest $1 \%$ displacements are plotted as red arrows.
}
\label{sfig3}
\end{figure*}

\noindent
{\bf Visualization of vibrational modes.-} 
The Letter also includes a detailed analysis of the nature of vibrational modes across the amorphisation transition. In particular, the degree of localization has been discussed in terms of participation ratios (see Fig.5). It is also instructive to directly visualize  a few eigen-vectors. We plot in Fig. \ref{sfig3} the displacement fields corresponding to the indicated eigen-frequencies, for $\lambda=1.0,\ 0.9,\ 0.81,$ and $0.8$.
We have considered a system of size $L=10a$ with $N=4,000$ particles.
At each $\lambda$, we plot typical vibrational modes extracted from the low and high frequency parts of the spectrum, respectively. In the case of localized modes with a participation ratio $PR_k < 0.1$, the largest $1 \%$ displacements are plotted as red arrows. In the perfect crystal case, $\lambda=1.0$, vibrational modes are extended and delocalized in the entire spectrum, as shown in Fig. \ref{sfig3}(a) and (b). As $\lambda$ decreases, high frequency vibrational modes become increasingly localized (see Fig. 5), as it is evident for the case $\lambda=0.9$, in Fig. \ref{sfig3}(d).  Below $\lambda=0.82$, localization also occurs at low frequencies (Fig. \ref{sfig3}(f-i)).

%%%%%%%%%%%%%%%%%%%%%%%% Supplementary Figure 3 %%%%%%%%%
%%%%%%%%%%%%%%%%%%%%%%%%%%%%%%%%%%%%%%%%%%%%%%%%%%%%%%%%%%%%%%%%%%%%%%%

%%%%%%%%%%%%%%%%%%%%%%%%%%%%%%%%%%%%%%%%%%%%%%%%%%%%%%%%%%%%%%%%%%%%%%%%%%%%%%%%%%%%%%%%%%%%%%%%%%%%%%%%%%%%%%%%%%%%%%%%%
\vspace{1.cm}

\noindent
{\bf Quantum corrections for thermal conductivity.-}
Classical molecular dynamics simulations cannot obviously grasp the temperature dependence of heat capacity, which is a genuine quantum effect. However, it is possible to qualitatively incorporate quantum corrections to the study of thermal conductivity (see, among others, Ref.~\cite{macgaughey_2006}).
In what follows all quantities are expressed in ``Argon" units, $\sigma =3.405\mathrm{\AA}$, $\epsilon/k_B=125.2{K}$, and $\tau = 2.11\text{ps}$. The ``classical" values of temperature, $T_{\text{MD}}$ and thermal conductivity, $\kappa_{\text{MD}}$, which are obtained from molecular dynamics simulation, can be mapped onto the ``real" values of $T_{\text{real}}$ and $\kappa_{\text{real}}$. The value of $T_{\text{MD}}$ is mapped onto $T_{\text{real}}$ by equating the energies of classical and quantum system:
\begin{equation}
\begin{aligned}
k_B T_{\text{MD}} &= \int \hbar \omega \left( \frac{1}{2} + \frac{1}{\exp(\hbar \omega/k_B T_{\text{real}})-1} \right) g(\omega) d\omega.
\end{aligned} 
\label{qc1}
\end{equation}
Here, $\hbar = h/2\pi$, $h$ is the Planck constant, and $g(\omega)$ is the vibrational density of state, which is shown in Fig. 3(a). The first term of right-hand side in Eq. (\ref{qc1}) is the ``zero point energy" of the quantum system. In the present study, we have not taken into account the zero point energy.

%%%%%%%%%%%%%%%%%%%%%%%% Supplementary Figure4 %%%%%%%%%
\begin{figure}[t]
\begin{center}
\includegraphics[scale=1]{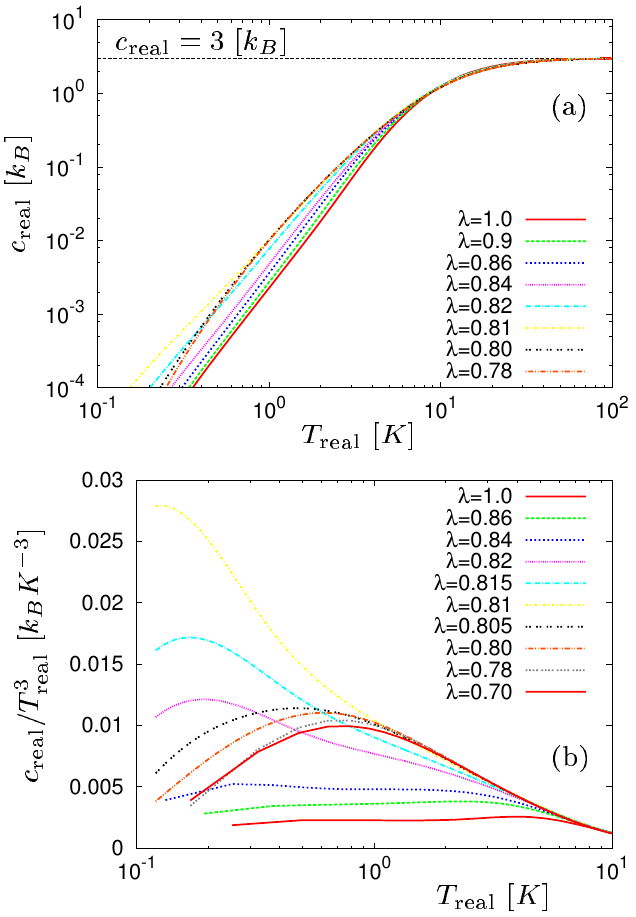}
\end{center}
\vspace*{-8mm}
\caption{The temperature $T_\text{real}$ dependence of (a) $c_\text{real}$ and (b) $c_\text{real}/T^3_\text{real}$.
}
\label{sfig4}
\end{figure}
%%%%%%%%%%%%%%%%%%%%%%%%%%%%%%%%%%%%%%%%%%%%%%%%%%%%%%%%%%%%%%%%%%%%%%%%%%%%%%%%%%%%%%%%%%%%%%%%%%%%%%%%%%%%%%%%%%%%%%%%%

Mapping of $\kappa_{\text{MD}}$ onto $\kappa_{\text{real}}$ is expressed by
\begin{equation}
\begin{aligned}
\kappa_{\text{real}} = \frac{C_{\text{real}}}{C_{\text{MD}}} \kappa_{\text{MD}},
\end{aligned} \label{qc2}
\end{equation}
where $C_{\text{MD}}$ and $C_{\text{real}}$ are the heat capacities of classical and quantum systems, respectively. The value $C_{\text{real}}$ is calculated as
\begin{equation}
\begin{aligned}
{c_\text{real}} &= \frac{C_\text{real}}{N-1} \\
&= 3 k_B \int \frac{ (\hbar \omega/k_B T_\text{real})^2 \exp(\hbar \omega/k_B T_{\text{real}})}{(\exp(\hbar \omega/k_B T_{\text{real}})-1)^2} g(\omega) d\omega,
\end{aligned} \label{spheat}
\end{equation}
where $c_\text{real}$ is the heat capacity per particle.
In Fig. \ref{sfig4}(a), we show $c_\text{real}$ as a function of $T_\text{real}$.
As $T_\text{real}$ increases, $c_\text{real}$ tends to the classical value $3 k_B$.
For the perfect crystal case, $\lambda =1.0$, $c_\text{real} \sim T^3_\text{real}$ in the low temperature regime, e.g., the Debye-model prediction holds.
However, as $\lambda$ decreases, the temperature dependence of $c_\text{real}$ deviates from the Debye model. We plot $c_\text{real}/T_\text{real}^3$ in Fig. \ref{sfig4} (b).
We clearly see that the peak value in $c_\text{real}/T_\text{real}^3$ is pronounced at the transition point $\lambda=\lambda^\ast \simeq 0.81$. This effect originates from the excess of low frequency modes in $g(\omega)/\omega^2$, as shown in Fig.~3(b).

%%%%%%%%%%%%%%%%%%%%%%%%%%%%%%%%%%%%%%%%%%%%%%%%%%%%%%%%%%%%%%%%%%%%%%%%%%%%%%%%%%%%%%%%%%%%%%%%%%%%%%%%%%%%%%%%%%
\begin{acknowledgements}
We acknowledge useful discussions with Dr. O.~N. Bedoya-Martinez.
This work was supported by the Nanosciences Foundation of Grenoble.
JLB is supported by the Institut Universitaire de France. 
\end{acknowledgements}


\begin{thebibliography}{1}
\bibitem{elliott_1984}
\Name{S.R. Elliott}
\Book{Physics of amorphous materials}
\Publ{Wiley, New York}
\Year{1984}.

\bibitem{phillips_1981}
\Name{W.A. Phillips}
\Book{Amorphous Solids: Low Temperature Properties}
\Publ{Springer, Berlin}
\Year{1981}.

\bibitem{Klinger_1983}
\Name{M.I. Klinger}
\REVIEW{Phys. Rep.}{94}{1983}{183}. 

\bibitem{schirmacher_1998}
\Name{W. Schirmacher \textit{et al.}}
\REVIEW{Phys. Rev. Lett.}{81}{1998}{136}. 

\bibitem{grigera_2003}
\Name{T. S. Grigera \textit{et al.}}
\REVIEW{Nature}{422}{2003}{289}. 

\bibitem{Wyart}
\Name{M. Wyart}
\Review{Annales de Physiques}{30}{2005}{1}

\bibitem{Mermet}
\Name{E. Duval and A. Mermet}
\Review{Phys. Rev. B} {58} {1998} {8159}

\bibitem{schirmacher_2006}
\Name{W. Schirmacher}
\REVIEW{Europhys. Lett.}{73}{2006}{892}. 

\bibitem{schirmacher_2007}
\Name{W. Schirmacher \textit{et al.}}
\REVIEW{Phys. Rev. Lett.}{98}{2007}{025501}. 

\bibitem{yoshimoto_2004}
\Name{K. Yoshimoto \textit{et al.}}
\REVIEW{Phys. Rev. Lett.}{93}{2004}{175501}. 

\bibitem{tsamados_2009}
\Name{M. Tsamados \textit{et al.}}
\REVIEW{Phys. Rev. E}{80}{2009}{026112}. 

\bibitem{makke_2011}
\Name{A. Makke \textit{et al.}}
\REVIEW{Macromol. Theory Simul.}{20}{2011}{826}. 

\bibitem{leonforte_2005}
\Name{F. Leonforte \textit{et al.}}
\REVIEW{Phys. Rev. B}{72}{2005}{224206}. 

\bibitem{marruzzo_2013}
\Name{A. Marruzzo \textit{et al.}}
\REVIEW{Nature Scientific Reports}{3}{2013}{1407}. 

\bibitem{Bocquet_1992}
\Name{L. Bocquet \textit{et al.}}
\REVIEW{J. Phys.: Condens. Matter}{4}{1992}{2375}. 

\bibitem{bernu_1987}
\Name{B. Bernu \textit{et al.}}
\REVIEW{Phys. Rev. A}{36}{1987}{4891}. 

\bibitem{Mizuno_2013}
\Name{H. Mizuno \textit{et al.}}
\REVIEW{Phys. Rev. E}{87}{2013}{042306}. 

\bibitem{lutsko_1988}
\Name{J. F. Lutsko}
\REVIEW{J. Appl. Phys.}{64}{1988}{1152}. 

\bibitem{supplement}
{http://arxiv.org/abs/1308.5135}

\bibitem{tanguy_2002}
\Name{A. Tanguy \textit{et al.}}
\REVIEW{Phys. Rev. B}{66}{2002}{174205}. 

\bibitem{Jasiukiewicz_2003}
\Name{Cz. Jasiukiewicz and V. Karpus}
\REVIEW{Solid State Commun.}{128}{2003}{167}. 

\bibitem{monaco_2009}
\Name{G. Monaco and S. Mossa}
\REVIEW{Proc. Natl. Acad. Sci. USA}{106}{2009}{16907}.

\bibitem{mazzacurati_1996}
\Name{V. Mazzacurati \textit{et al.}}
\REVIEW{Europhys. Lett.}{34}{1996}{681}. 

\bibitem{shober_2004}
\Name{H. R. Schober and G. Ruocco}
\REVIEW{Philos. Mag.}{84}{2004}{1361}. 

\bibitem{shintani_2008}
\Name{H. Shintani and H. Tanaka}
\REVIEW{Nature Mater.}{7}{2008}{870}. 

\bibitem{tan_2012}
\Name{P. Tan \textit{et al.}}
\REVIEW{Phys. Rev. Lett.}{108}{2012}{095501}. 

\bibitem{macgaughey_2006}
\Name{A. J. H. McGaughey and M. Kaviany}
\Book{Advances in Heat Transfer}
\Editor{G. A. Greene, Y. I. Cho, J. P. Hartnett, and A. Bar-Cohen}
\Publ{Elsevier, New York}
\Year{2006}
\Vol{39}
\Page{169-255}.

\bibitem{masciovecchio_2006}
\Name{C. Masciovecchio \textit{et al.}}
\REVIEW{Phys. Rev. Lett.}{97}{2006}{035501}. 

\bibitem{allen_1993}
\Name{P. B. Allen and J.L. Feldman}
\REVIEW{Phys. Rev. B}{48}{1993}{12581} .

\bibitem{christensen_1975}
\Name{D. K. Christensen and G.L. Pollack}
\REVIEW{Phys. Rev. B}{12}{1975}{3380}. 

\bibitem{cahill_1992}
\Name{D. G. Cahill \textit{et al.}}
\REVIEW{Phys. Rev. B}{46}{1992}{6131}. 

\bibitem{silbert_2005}
\Name{L. E. Silbert \textit{et al.}}
\REVIEW{Phus. Rev. Lett.}{95}{2005}{098301}. 

\bibitem{turney_2009}
The quantum correction  we applied  is a global one, which does not take into account the mode-dependent occupation discussed in 
\Name{J. E. Turney \textit{et al.}}
\REVIEW{Phys. Rev. B}{79}{2009}{224305}. 

\bibitem{xu_2009}
\Name{N. Xu \textit{et al.}}
\REVIEW{Phys. Rev. Lett.}{102}{2009}{038001}. 

\bibitem{ma_2013}
\Name{J. Ma \textit{et al.}}
\REVIEW{Nature Nanotech.}{8}{2013}{445}. 

\end{thebibliography}
\end{document}